\documentclass[twocolumn,prl,showpacs,showkeys,superscriptaddress]{revtex4-1}

\usepackage{amsmath}
\usepackage[pdftex]{graphicx}
\DeclareGraphicsExtensions{.pdf}
\usepackage{subfig}
\usepackage{bm}
\usepackage{color}
\usepackage{paralist}

\newcommand\gonetau{\ensuremath{g^{\left(1\right)}(\tau)}}

\begin{document}

\title{Build up of off-diagonal long-range order in microcavity
exciton-polaritons  across the parametric threshold}

\author{R. Spano}
\email{rita.spano@uam.es}
\affiliation{Dept. F\'isica Materiales,Universidad Autonoma de Madrid, Madrid 28049, Spain}

\author{J. Cuadra}
\affiliation{Dept. F\'isica Materiales,Universidad Autonoma de Madrid, Madrid 28049, Spain}

\author{C. Lingg}
\affiliation{Dept. F\'isica Materiales,Universidad Autonoma de Madrid, Madrid 28049, Spain}

\author{D. Sanvitto}
\altaffiliation[Current address: ]{NNL, Istituto Nanoscienze - CNR, Via Arnesano, 73100 Lecce, Italy}
\affiliation{Dept. F\'isica Materiales,Universidad Autonoma de Madrid, Madrid 28049, Spain}

\author{M.D. Martin}
\affiliation{Dept. F\'isica Materiales,Universidad Autonoma de Madrid, Madrid 28049, Spain}

\author{P.R. Eastham}
\affiliation{School of Physics, Trinity College, Dublin 2, Ireland}

\author{M. van der Poel}
\affiliation{DTU Fotonik, Tech. Univ. Denmark, \O rsteds Plads 343 DK-2800 Kgs. Lyngby, Denmark}

\author{J. M. Hvam}
\affiliation{DTU Fotonik, Tech. Univ. Denmark, \O rsteds Plads 343 DK-2800 Kgs. Lyngby, Denmark}

\author{L. Vi{\~n}a}
\affiliation{Dept. F\'isica Materiales,Universidad Autonoma de Madrid, Madrid 28049, Spain}

\begin{abstract}

  We report an experimental study of the spontaneous spatial and temporal coherence of polariton condensates generated in the optical parametric oscillator configuration, below and at the parametric threshold, and as a function of condensate area. Above the threshold we obtain very long coherence times (up to $3$ ns) and a spatial coherence extending over the entire condensate ($40 \, \mu\mathrm{m}$). The very long coherence time and its dependence on condensate area and pump power reflect the suppression of polariton-polariton interactions by an effect equivalent to motional narrowing.
\end{abstract}



\maketitle

Planar microcavity exciton-polaritons are composite particles
generated by strong coupling between excitons and cavity photons that
follow the Bose statistics. They have been extensively investigated in
the past decades for their unique properties, such as very light
effective mass and the consequent high critical temperature for
condensation \cite{Deveaud}. It is well known that the occurrence of
such a phase transition is accompanied by the onset of long range
phase coherence \cite{Penrose}. In fact, since the unambiguous demonstration of polariton condensation its coherence properties have been
investigated \cite{Kasprzak}. Moreover, the extension and eventually
the decay of the spatial coherence can give useful information on the
kind of transition the system undergoes
\cite{Pitaevskii,Laussy}. Recently, polariton condensates have
attracted intense attention for their potential application in the
field of quantum information, and also some properties suitable for
such applications, such as Josephson oscillations and spin-switching,
have been already demonstrated \cite{Lagoudakis,Amo}. Therefore the
achievement of extended spatial and temporal coherence for the
polariton condensate is crucial.

Here we study the spatial and temporal coherence properties of
nonequilibrium polariton condensates under an optical parametric
oscillator (OPO) regime. The OPO is a third order non-linear
phenomenon, arising from interactions between the excitonic components
of the polaritons. It takes place when two pump polaritons at the
inflection point of the lower polariton branch (LPB), scatter
efficiently into a signal and an idler polariton
\cite{Stevenson,Baumberg}. Spatial coherence properties have been
studied theoretically for the OPO condensate by Carusotto and Ciuti
\cite{Carusotto}. They investigated numerically the first order
coherence function ($g^{\left(1\right)}$) for a finite condensate
paying special attention to its behaviour across the OPO parametric
threshold ($E_{Th}$) \cite{note}. It is found that, for excitation
frequencies $\omega_p$ below that of the threshold,
$g^{\left(1\right)}$ has a finite correlation length. Increasing
$\omega_p$ in order to approach $E_{Th}$, mantaining fixed the pump
angle (and therefore the wavevector), they predict the build up of
macroscopic phase coherence extending over the entire condensate
\cite{Carusotto}. In this regime the spatial fluctuations are
negligible, so the temporal coherence properties should be captured by
the theory of Whittaker and Eastham \cite{Whittaker}. In this theory
the temporal coherence is limited by fluctuations in the particle
number, which due to the polariton-polariton interactions imply a
broadening of the emission
\cite{Yamamoto,Porras,KrizhanovskiiSanvitto}. This broadening
mechanism, however, could be suppressed by motional narrowing
\cite{Whittaker}, so that for appropriate pump powers and condensate
areas very long coherence times could be obtained.  Despite the fact
that long coherence times and extended spatial coherence have been
predicted theoretically for the OPO configuration, only now our
experiments confirm such predictions.

\begin{figure}
\includegraphics[width=0.49\textwidth]{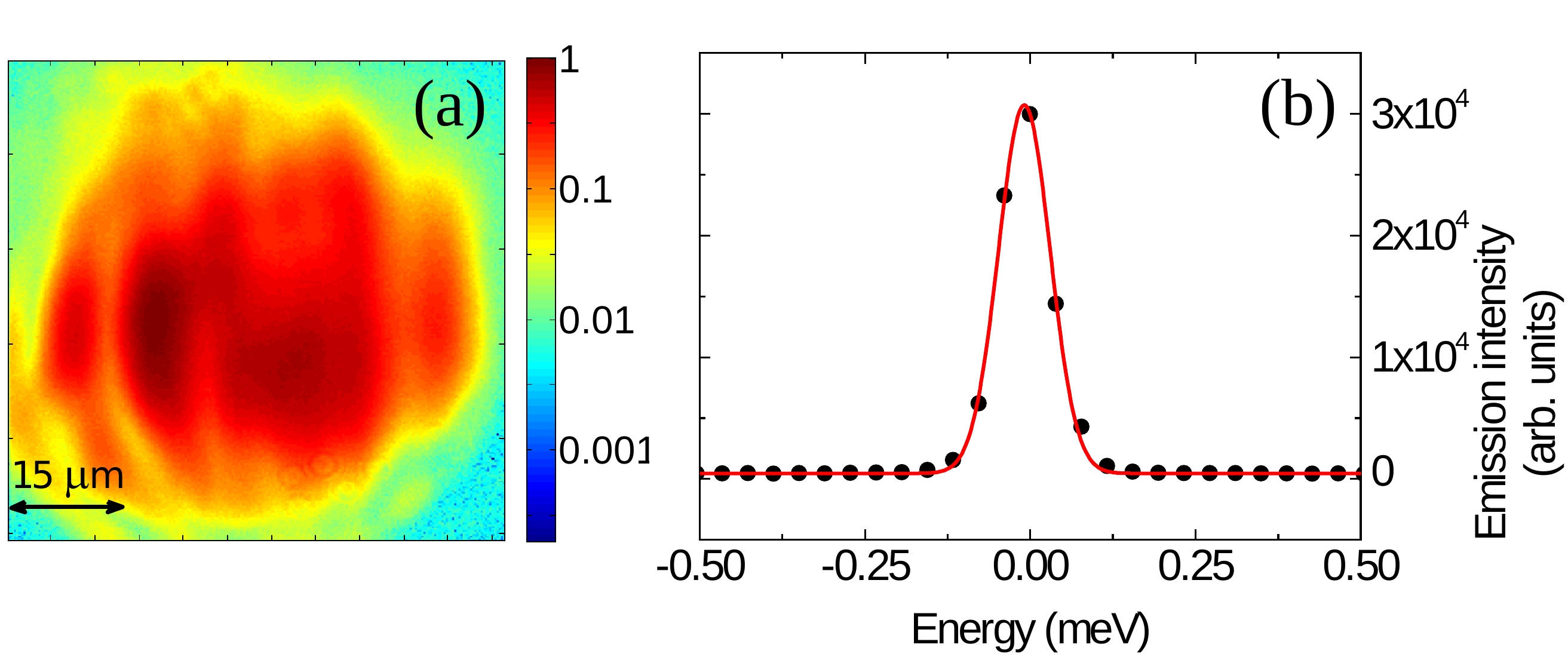}
\caption{(color online) (a) Typical real space emission intensity (color scale) from
  the signal condensate generated by parametric scattering, at
  detuning $\delta\sim -3.8$ meV and pump power $P=10 \,
  \mathrm{kW}/\mathrm{cm}^2$. (b) Corresponding energy spectrum of the
  signal emission (spectrometer limited).}
 \label{fig:fig1}
\end{figure}

One of the main factors that limits the coherence is the quality of
the cavity: the presence of defects creates a disorder potential that
traps the condensate, potentially leading to multi-mode and
inhomogeneous states \cite{Krizhanovskii}. Another, very important,
detrimental effect is caused by the fluctuations of the excitation
laser that hinder the attainment of the intrinsic coherence of the
condensate. Many studies \cite{Kasprzak,Krizhanovskii,Deng} have been
performed by non-resonantly pumping the microcavity, in such cases the
resulting distribution of the population at the bottom of the lower
polariton branch is subjected to fluctuations due to the reservoir of
particles at the bottleneck. These fluctuations, broadening the
distribution of polaritons in energy and momentum space, translate,
according to the Wiener-Khinchin identity \cite{Baltes,Roussignol},
into a faster decay of the temporal and spatial coherence.  Although
analogous broadening by the fluctuating population of pump polaritons
can occur in the OPO \cite{KrizhanovskiiSanvitto}, the threshold pump
density ($P_{Th}$) for a resonant-gain process is much lower than that
for non-resonant gain and the effect of the reservoir polaritons is
less important. Thus the coherence exhibited by an OPO
polariton-condensate is expected to decay over much longer times and
larger distances than the one produced by non-resonant techniques.  To
avoid that the pump fluctuations conceal the genuine coherence
properties of the condensate we use a CW monomode laser with a very
narrow bandwidth of $75$ kHz to excite the sample, which is a high-quality $\lambda$-microcavity grown by molecular beam epitaxy,
composed by $16$ periods of $\lambda/4$
$\text{AlAs}/\text{Al}_{0.15}\text{Ga}_{0.85}\text{As}$ on its top and
$25$ periods on its bottom. A single $10$ nm
$\text{GaAs}/\text{Al}_{0.3}\text{Ga}_{0.7}\text{As}$ QW is placed at
the antinode of the cavity. The Rabi splitting is $4.2$ meV and we select a
detuning of $\delta\sim-3.8$ meV. In order to investigate the spatial
coherence properties of a true $2\text{D}$ condensate we create one
with a diameter of $\sim40 \, \mu\text{m}$, by pumping resonantly the
LPB close to its inflection point, with a Gaussian laser beam,
maintaining the temperature of the sample at $10$ K. We adopt the
experimental conditions to drive the sample into the OPO regime at the
parametric threshold, $E_{Th}$, and then we follow a similar approach
to that in \cite{Carusotto}, i.e., we decrease the pump frequency
$\omega_p$ keeping fixed the angle of incidence. Typical results for
the OPO signal are shown in Fig. \ref{fig:fig1}. They reveal a bright
and rather homogeneous emission in real space
(Fig. \ref{fig:fig1}(a)), with a spectrometer-limited emission
spectrum (Fig.  \ref{fig:fig1}(b)).

Therefore, to evaluate the coherence of the condensate, since its true
energy linewidth is masked by our spectrometer, the signal emission
is analyzed with the help of a Mach-Zehnder interferometer. On one of
the arms of the interferometer a retro reflector mirror is mounted,
which flips the image of the condensate in a centrosymmetric
way. Recombining, at the output of the interferometer, the image of the
condensate with its flipped image allows evaluating the coherence of
each point $\left(x, y\right)$ with respect to its opposite
counterpart $\left(-x, -y\right)$. In this way we access to the first
order coherence function $g^{\left(1\right)}\left[\left(x,
    y\right),\left(-x,-y\right)\right]=\frac{\left\langle
    E^{\ast}\left(x,
      y\right)E\left(-x,-y\right)\right\rangle}{\left\langle
      E^{\ast}\left(x, y\right)\right\rangle \left\langle
      E\left(-x,-y\right)\right\rangle}$ \cite{Penrose,Kasprzak}. To
extract the coherence of the condensate the retroreflector mirror is
moved, using a piezo translation stage, by a few wavelengths around
the zero delay position, adding a short delay $\Delta{t}$ to the
retroreflector arm with respect to the other arm. For each
measurement the intensities of the two arms were acquired and used in
the following normalized expression of the interference pattern:
$I_{Norm}\left[\left(x,
    y\right),\left(-x,-y\right)\right]=\frac{I_{Tot}-I_{RR}-I_{M}}{2
  \sqrt{I_{RR}I_{M}}} = g^{\left(1\right)}\left[\left(x,
    y\right),\left(-x,-y\right)\right]\sin\left(\Delta{t} +
  \varphi_0\right)$ where $I_{Tot}$ is the total intensity of the
interference pattern, $I_{RR}$ ($I_{M}$) is the intensity recorded
from the arm with the retroreflector (single mirror), and $\varphi_0$
is the initial phase of the condensate. For each point of the
condensate a sinusoidal evolution of the normalized interference as a
function of the delay $\Delta{t}$ is recorded. The first order
coherence of each point of the condensate is extracted with the help
of the previous formula; and repeating this procedure for each point
we can reconstruct the entire coherence map.

\begin{figure*}
\includegraphics[width=0.80\textwidth]{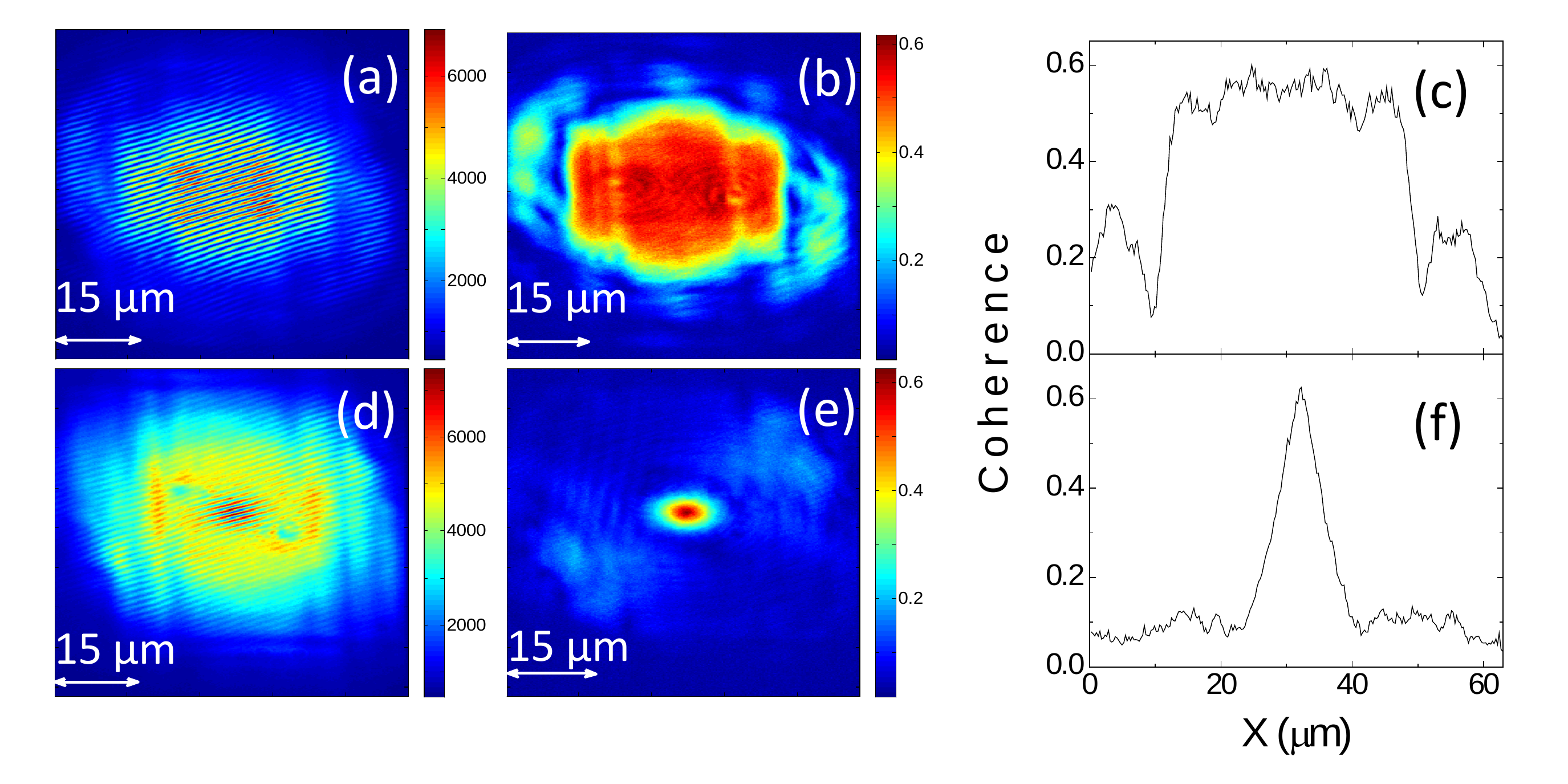}
\caption{(color online) Interference pattern (a/d) and corresponding coherence (b/e) of
  the condensate generated at/below the parametric threshold, $E_{Th}$,
  corresponding to a pump energy $E_1=1552.57$ meV/ $E_2=1552.56$ meV and power density
  $P=15.8 \, \mathrm{kW}/\mathrm{cm}^2$. Horizontal profiles at the
  center (c/f) showing the constant coherence along the entire
  condensate for $E_1$, and an exponentially decaying one for $E_2$.}
 \label{fig:fig2}
\end{figure*}
		 At $E_{Th}$, the interference pattern extends along
the entire condensate (Fig. \ref{fig:fig2}(a)) and the corresponding
coherence map (Fig. \ref{fig:fig2}(b)) reveals that a constant phase
is developed through its entire area, validating the theoretical predictions
\cite{Carusotto}. This is shown in detail in Fig. \ref{fig:fig2}(c), which depicts the coherence of a horizontal profile ($0.26 \, \mu\mathrm{m}$
width) taken at the center of the map.  The fact that the degree of
coherence is almost flat along the entire condensate demonstrates that
only a single coherent mode is formed, in contrast with what has been
found in previous works \cite{Krizhanovskii,Love} where disorder
of the sample caused the development of several modes. The extension of
the coherence here is $\sim40 \, \mu\text{m}$, the largest reported
until now in $2$D microcavity polaritons, to the best of our
knowledge; only in 1D microcavities, for propagating
condensates, a larger spatial coherence has
been shown \cite{Bloch}. It is worth to note that in an OPO process, the coherence
properties of the signal are not inherited from the excitation laser
\cite{Wouters}. Although the theory of an infinite $2$D condensate
predicts the absence of long-range order, and a power-law decay of the
spatial coherence \cite{Pitaevskii}, in a
finite dimension system such as ours, determined by the pump laser
size, a constant coherence of the condensate can be obtained
\cite{Kasprzak}.

By decreasing the pump energy by $\delta E=0.01$ meV towards lower
energies, leaving fixed the pump angle, the phase matching conditions
for the OPO are not fulfilled anymore, although an emission is
obtained at the bottom of the LPB. The interferometric analysis of the
emission from the cloud of non-condensed polaritons gives an
interference pattern only at the center of the emission (Fig.\ref{fig:fig2}(d)). The corresponding coherence map
(Fig.\ref{fig:fig2}(e)) reveals a $\sim7 \, \mu\mathrm{m}$ coherence
length, of a few polariton de Broglie wavelengths, extracted by
fitting the profile of the spatial decay
(Fig.\ref{fig:fig2}(f)). These results demonstrate the predictions of
Ref.  \cite{Carusotto} about the coherence of polariton emission below
and at the condensation threshold.

\begin{figure}
  \includegraphics[width=0.50\textwidth]{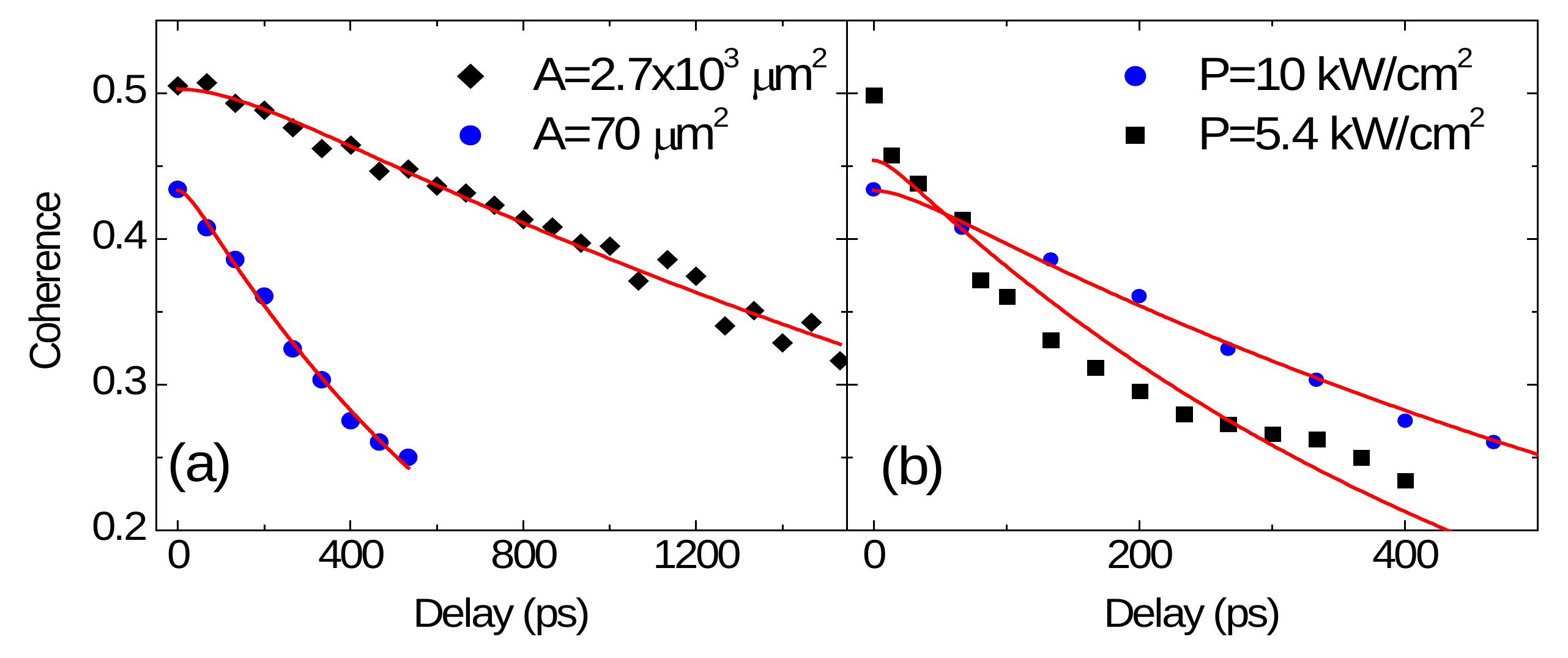}
  {\caption{(color online) (a)Temporal coherence decay above parametric threshold for
      condensate areas $A=70 \mu\mathrm{m}^2$ (blue dots) and
      $2.7\times10^3 \mu\mathrm{m}^2$ (black diamonds) at a pump power
      density $P=10 \mathrm{kW}/\mathrm{cm}^2$; (b) Temporal coherence
      decay for $P=5.4 \mathrm{kW}/\mathrm{cm}^2$ (black squares) and $10
      \mathrm{kW}/\mathrm{cm}^2$ (blue dots) at $A=70
      \mu\mathrm{m}^2$. Lines are fits to Eq.($1$).}
 \label{fig:fig3}}
\end{figure}

Measuring $g^{\left(1\right)}$ at a given point for different time
delays $\tau$ permits the evaluation of the signal temporal coherence
$g^{\left(1\right)}\left(\tau\right)=\frac{\left\langle
    E^{\ast}\left(t\right)E\left(t+\tau\right)\right\rangle}{\left\langle
      E^{\ast}\left(t\right)\right\rangle \left\langle
      E\left(t\right)\right\rangle}$. In our case we obtain
$g^{\left(1\right)}$ for different values of the condensate area
$A$ and pump power density $P$, as shown in Fig. \ref{fig:fig3}. We obtain a predominantly
exponential decay for $g^{\left(1\right)}\left(\tau\right)$, with a
coherence time reaching $T_c=(3.2\pm0.8)$ ns. This is much longer than
the largest value reported in the literature until now for the same
material system \cite{KrizhanovskiiSanvitto}.

Despite the lack of spatial fluctuations, the coherence time of our
condensate, though long, is clearly finite. In general, such a finite
correlation time in a state with perfect spatial order is caused by
finite-size fluctuations \cite{EasthamFiniteSize}, and reflects the
absence of true phase transitions in finite systems. A well-known
example is the Schawlow-Townes form for the coherence time of a single
laser mode with an average of $\bar{N}$ photons, $T_c\propto
\bar{N}$. Since $\bar{N}\propto A$ when the control parameter $P$ is
constant, the Schawlow-Townes result implies the scaling form
$T_c\propto A$. However, as shown in detail below, this scaling law is
violated by more than an order of magnitude for our condensate, and
our results cannot be understood as a straightforward effect of
increasing the particle number.  Our results may be interpreted in
terms of the theoretical model of Ref. \cite{Whittaker}, in which the
linewidth arises from fluctuations in the number of particles. Due to
the interactions, such number fluctuations imply energetic fluctuations
of the emission, leading to a broadened line or a decay of
$g^{\left(1\right)}\left(\tau\right)$. This is captured by Kubo's
result
\begin{equation}
  \left|g^{\left(1\right)}\left(\tau\right)\right|=\exp\left[-\frac{2
      \tau_r^2}{\tau_c^2}\left(e^{-\tau/\tau_r}+\frac{\tau}{\tau_r}-1\right)\right], \label{eq:kubo} \end{equation}
for the emission from a transition whose energy fluctuates, where
$\tau_c$ is determined by the width of the (generically Gaussian) distribution of the
energy level and $\tau_r$ is the characteristic timescale on
which the fluctuations occur. If $\tau_r \gg \tau_c$ the fluctuations
are slow and the emission reflects the energy distribution of the
level, so that the coherence decays on timescale $\tau_c$. If however
$\tau_r \ll \tau_c$ a motional narrowing effect leads to an
exponential decay, with a much longer coherence time
$T_c=\tau_c^2/2\tau_r$.

To elucidate the effects of number fluctuations, we fit the data in
Fig.\ \ref{fig:fig3} to the Kubo formula (\ref{eq:kubo}). Pumping with
the same power density $P = 10 \, \mathrm{kW}/\mathrm{cm}^2$, the
condensate with a larger area $A=2.7\times10^3 \mu\mathrm{m}^2$
presents $\tau_c=(0.97\pm0.11)$ ns and $\tau_r=(0.15\pm0.02)$ ns
fulfilling the inequality for the motional narrowing regime and giving
a coherence time $T_c=(3.2\pm0.8)$ ns. With a smaller area $A=70
\mu\mathrm{m}^2$, we find a decrease of both parameters $\tau_c=(0.20\pm0.08)$ ns and $\tau_r=(0.023\pm0.007)$ ns, still
indicating the system being in the motional narrowing regime but
giving a shorter $T_c=(0.9\pm0.4)$ ns. As well as moving to
larger areas, another way to have longer decay times suggested by
\cite{Whittaker} is to increase the pump power. The results shown in
Fig. \ref{fig:fig3}(b), corresponding to two different powers keeping
now the size of the condensate constant, also demonstrate that
increasing the power a longer coherence time is achieved. In fact the
coherence time for a condensate of relatively small area $A=70\,
\mu\mathrm{m}^2$ goes from $T_c=(0.5\pm0.3)$ ns at $P=5.4 \,
\mathrm{kW}/\mathrm{cm}^2$, to $T_c=(0.9\pm0.4)$ ns at $P=10 \,
\mathrm{kW}/\mathrm{cm}^2$. We note that the coherence decay of the
smallest condensate at lowest power, black squares in
Fig. \ref{fig:fig3}(b), has considerable structure and does not follow
the Kubo form, suggesting that this condensate is not a single mode.
From the Kubo fits to our data, for a given power, we can
extract the scaling $\tau_c \propto A^{0.4\pm 0.1}$. This is
consistent with the dominant dephasing mechanism being the
polariton-polariton interaction, suppressed by motional narrowing, in
which case the Kubo formula should hold with $\tau_c \propto A^{0.5}$
\cite{Whittaker}.  
Although this implies an increase in the coherence time, $T_c$, with
area, our results show that this increase is restrained because
$\tau_r$ also increases, so that motional narrowing becomes less
effective.  

The relatively small size of our system and the use of pump powers
which are close to threshold, $P_{Th}$, are responsible for the
observed increase of $\tau_r$ with area. In the thermodynamic limit,
$A\to\infty$, the occupation, $n$, of a single mode with linear
gain $\gamma$, linear loss $\gamma_c$ and nonlinear gain $\Gamma$ obeys the mean-field rate
equation $dn/dt=(\gamma -\gamma_c - \Gamma n)n$ \cite{PREastham}. Linearizing we see that the
damping time for number fluctuations is $\tau_r=1/|\gamma-\gamma_c|$. This
timescale is therefore independent of area when the rate equation is
valid. However, at power threshold $\gamma=\gamma_c$ and the rate equation
predicts a divergence in the relaxation time, which in the finite
system must be cut off by fluctuations. Thus, in the threshold region,
$\tau_r$ initially grows with area, as we observe, before eventually
saturating. We obtain a similar behavior from numerical solutions to
the dynamical model described in Ref. \cite{Whittaker} in the
threshold region. If we assume a polariton lifetime $\tau_0=3\;
\mathrm{ps}$, then our results for $P = 10 \,
\mathrm{kW}/\mathrm{cm}^2$, $A= 70\, \mu\mathrm{m}^2$ imply the
critical slowing factor $\tau_r/\tau_0\approx 8$. Fitting this to the
dynamical model at threshold gives a resonable value for the gain
saturation density, $\rho_s=n_s/A=230/(70\mu
\mathrm{m^2})$. Since $\rho_s$ is independent of $A$, we use this value in the model to predict 
$\tau_r= 140$ ps for our largest condensate,
consistent with experiments.

For completeness we also studied the temporal coherence (not shown)
under two further conditions: a) Exciting the LPB at the inflection
point, for power densities slightly above but very close to $P_{Th}$, we find a Gaussian decay of \gonetau, as
previously observed for condensates with non-resonant pumping
\cite{Krizhanovskii}. This is consistent with the expected increase in
$\tau_r$ approaching $P_{Th}$ causing a crossover from the
motional-narrowing to the static regime. b) Exciting far from the
inflection point (no condensation achieved) we observe a very fast
exponential decay of coherence on the range of the polariton lifetime.

In summary, we have investigated the spatial and temporal coherence
properties of polariton condensates generated by parametric
scattering. The spatial coherence was studied below and above the OPO
condensation threshold, obtaining the largest value for the coherence
length measured for $2$D GaAs microcavity polaritons. Measurements of
the temporal coherence reveal a predominantly exponential decay caused
by polariton-polariton interactions in a motional narrowing
regime. Although similar to the exponential decay associated with the
Schawlow-Townes linewidth, the mechanism is completely different, and
gives a different dependence of coherence time on condensate area. By
constructing condensates of large area we are able to achieve long
coherence times, which is a crucial step forward for exploiting
polariton condensates in quantum and ultrafast devices.

Acknowledgements: The work was supported by the FP7 ITNs Clermont4
(235114) and Spin-optronics (237252), Spanish MEC (MAT2011-22997), CAM
(S-2009/ESP-1503), and Science Foundation Ireland SIRG/I1592 (PRE). We
thank C. Tejedor and I. Carusotto for valuable discussions.


\begin{thebibliography}{10}

\bibitem{Deveaud} \textit{Physics of Semiconductor Microcavities}, ed. B. Deveuad (Wiley-VCH, Berlin, 2007).
\bibitem{Penrose} O. Penrose and L. Onsager, Phys. Rev. \textbf{104}, 576 (1956).
\bibitem{Kasprzak} J. Kasprzak \textit{et al.}, Nature \textbf{443}, 409 (2006).
\bibitem{Pitaevskii}L. Pitaevskii and S. Stringari, Bose-Einstein condensation (Oxford University Press, Oxford,
2003)
\bibitem{Laussy} F. P. Laussy \textit{et al.}, Phys. Rev. Lett. \textbf{93}, 016402 (2004).
\bibitem{Lagoudakis} K. G. Lagoudakis \textit{et al.}, Phys. Rev. Lett. \textbf{105}, 120403 (2010).
\bibitem{Amo} A. Amo \textit{et al.}, Nature Photonics \textbf{4}, 361 (2010).
\bibitem{Stevenson} R. M. Stevenson \textit{et al.}, Phys. Rev. Lett. \textbf{85}, 3680 (2000).
\bibitem{Baumberg} J. J. Baumberg \textit{et al.} Phys. Rev. B \textbf{62}, R16247 (2000).
\bibitem{Carusotto} I. Carusotto and C. Ciuti, Phys. Rev. B \textbf{72}, 125335 (2005).
\bibitem{note} In this work we consider two kinds of threshold,
  associated with tuning the pump energy and power. They are denoted
  $E_{Th}$ and $P_{Th}$ respectively.
\bibitem{Whittaker} D. M. Whittaker and P. R. Eastham, EPL \textbf{87}, 27002 (2009).
\bibitem{Yamamoto} F. Tassone and Y. Yamamoto, Phys. Rev. A \textbf{62}, 063809 (2000).
\bibitem{Porras} D. Porras and C. Tejedor, Phys. Rev. B \textbf{67}, 161310(R) (2003).
\bibitem{KrizhanovskiiSanvitto} D. N. Krizhanovskii \textit{et al.}, Phys. Rev. Lett. \textbf{97}, 097402 (2006).
\bibitem{Krizhanovskii} D. N. Krizhanovskii \textit{et al.}, Phys.Rev. B \textbf{80}, 045317 (2009). 
\bibitem{Deng} H. Deng \textit{et al.}, Phys. Rev. Lett. \textbf{99}, 126403 (2007).
\bibitem{Baltes} H. P. Baltes, Appl. Phys. \textbf{12}, 221 (1977).
\bibitem{Roussignol} M. Richard, M. Wouters and L.S. Dang, in \textit{Optical Generation and Control of Quantum Coherence in Semiconductor Nanostructures}, NanoScience and Technology \textbf{146}, Chap.11, eds. G. Slavcheva and P. Roussignol (Springer-Verlag Berlin 2010).

\bibitem{Love} A. P. D. Love \textit{et al.}, Phys. Rev. Lett. \textbf{101}, 067404 (2008). 
\bibitem{Bloch} E. Wertz \textit{et al.}, Nature Phys. \textbf{6}, 860 (2010).
\bibitem{Wouters}	M. Wouters and I. Carusotto, Phys. Rev. A  \textbf{76}, 043807 (2007).
\bibitem{EasthamFiniteSize} P. R. Eastham and P. B. Littlewood, Phys. Rev. B \textbf{73}, 085306 (2006). 
\bibitem{PREastham} P.R. Eastham, Phys. Rev. B \textbf{78}, 035319 (2008).

\end{thebibliography}
\end{document}